\pdfoutput=1
\documentclass[a4paper,12pt,abstract=true]{scrartcl}
\usepackage{authblk}
\usepackage{amsmath,amssymb}
\usepackage{bm}
\usepackage{graphicx,color}
\usepackage[hidelinks]{hyperref}
\usepackage{braket}
\usepackage{comment}
\usepackage{booktabs}
\usepackage{cite}

\begin{document}

\title{Control of emission interval and timing in triggered periodic superradiance}
\author[1]{Hideaki~Hara\thanks{Email: \texttt{hhara@okayama-u.ac.jp}}}
\author[1]{Riku~Omoto}
\author[1]{Noboru~Sasao}
\author[1]{Akihiro~Yoshimi}
\author[1,2]{Junseok~Han}
\author[1]{Yasutaka~Imai}
\author[1]{Koji~Yoshimura}
\author[1]{Motohiko~Yoshimura}
\author[1]{Yuki~Miyamoto\thanks{Email: \texttt{miyamo-y@cc.okayama-u.ac.jp}}}

\affil[1]{Research Institute for Interdisciplinary Science, Okayama University, Okayama, 700-8530, Japan}
\affil[2]{Department of Physics and Astronomy, Seoul National University, Seoul, Korea}

\date{\normalsize\today}

\maketitle

\begin{abstract}
To achieve more controllable development of coherence in solids, 
we investigated the effect of a trigger laser 
tuned to the superradiance transition wavelength 
on periodic superradiance observed in an Er:YSO crystal.
For period control, applying the trigger laser reduced 
both the superradiance period and its variance, 
demonstrating enhanced controllability of 
coherence development dynamics.
As the trigger laser power increased, 
both the period and the number of emitted superradiance photons 
decreased while maintaining a proportional relationship.
This behavior is explained by a reduced superradiance threshold 
under a constant excitation rate 
and is reproduced by numerical simulations 
based on the Maxwell-Bloch equations.
For timing control, we found that superradiance could be triggered 
even when the excitation laser alone was insufficient.
This enabled us to control the emission timing of superradiance 
using short trigger pulses 
and provided a device capable of generating superradiance 
at desired timing.
\end{abstract}

\section{Introduction}
\label{sec:introduction}

\begin{figure*}[t]
\begin{center}
      \includegraphics[width=15cm]{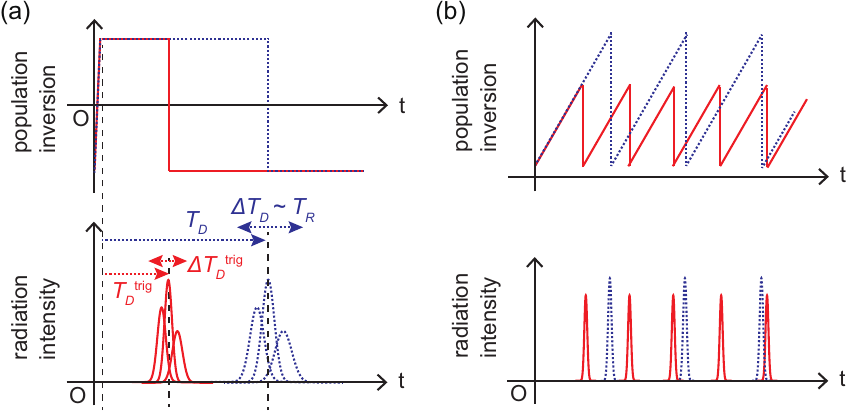}
       \caption{(a) Conceptual illustration of conventional 
       (non-periodic) SR.
       (Top) Temporal change in the population inversion 
       without (blue dotted line) and with (red solid line) 
       a trigger laser.
       The initial rapid increase represents excitation 
       by the pump pulse, 
       while the trigger laser---either CW or pulsed---serves as 
       a deterministic seed for coherence development.
       (Bottom) A single SR pulse is shown, 
       with pulse-to-pulse variation indicated.
       (b) Conceptual illustration of periodic SR.
       (Top) Temporal evolution of the population inversion 
       without (blue dotted line) and with (red solid line) 
       a trigger laser.
       Both the excitation and the trigger laser 
       are CW.
       (Bottom) Periodic SR pulses are shown.}
       \label{fig:conceptual}
\end{center}
\end{figure*}

Superradiance (SR) has been extensively studied 
\cite{SR-1954,SR-HFgas-1973,cascadeSR-Na-1976,SR-O2KCl-1982,SR-solid-review-2016,SR-nanocrystal-2018,SR-NVcenter-2018,An-singleatom-SR-2018,ErYSO-Padova-2020,ErYLF-Padova-2021,Kitano-SR-2024}, 
and a particular variant, triggered SR, has also attracted attention.
In the case of triggered SR, the emission timing of SR 
can be controlled by irradiating a laser at the same wavelength 
as the SR transition \cite{tipping-angle-1979}.
In general, SR exhibits a delay time ($T_{D}$) 
required for coherence to fully develop 
after the population inversion is formed 
as shown by the blue dotted line in Fig. \ref{fig:conceptual} (a).
In untriggered SR, 
coherence emerges from quantum fluctuations, 
where the electric field of spontaneous emission acts as a seed.
Consequently, $T_{D}$ reflects initial quantum fluctuation 
and exhibits significant pulse-to-pulse variation ($\Delta T_{D}$).
In triggered SR, by contrast, 
the external laser field---either continuous-wave 
(CW) or pulsed---serves as 
a deterministic seed for coherence development.
As a result, $T_{D}^{\mathrm{trig}}$, 
the delay time in the presence of a trigger laser, 
becomes shorter and more stable, 
with reduced shot-to-shot fluctuations, 
as shown by the red solid line in Fig. \ref{fig:conceptual} (a).
This can be regarded as a clear example of 
emission timing control in SR.
Such timing-controlled SR holds potential for applications 
in quantum information processing, 
such as utilizing deexcitation signals 
for quantum memory readout \cite{triggeredSR-spin-2023}.
Triggered SR also enables strong amplification 
induced by a trigger laser acting 
as a weak external input \cite{triggeredSR-1980,Kitano-SR-2024}.
Moreover, it has been proposed as a possible physical mechanism 
underlying astrophysical phenomena 
known as fast radio bursts (FRBs) \cite{triggeredSR-FRB-2018}.
Recently, triggered SR has been explored in Rydberg gases, 
where blackbody radiation induces emission \cite{SR-BBR-2023}.
These studies highlight a growing interest 
in externally controlling collective emission process 
for quantum technologies.

In the long history of research on SR, 
we recently observed quasiperiodic SR pulses in an Er:YSO crystal 
under CW laser excitation, 
without modulation of input parameters.
We referred to this phenomenon as 
``periodic superradiance" \cite{Hara-pSR}.
This behavior can be understood as a repeated cycle 
of population inversion buildup via pumping, 
followed by a sudden SR burst 
once the population inversion reaches a certain threshold, 
as shown in Fig. \ref{fig:conceptual} (b).
In this paper, we refer to SR behavior 
that is not periodic as ``conventional SR."
Unlike conventional SR, 
periodic SR exhibits regular, repeated pulses.
The emergence of periodic SR implies 
partial control of coherence development; 
indeed, the SR period varies with excitation laser power, 
though with considerable fluctuation.

Inspired by triggered SR, we hypothesize that 
applying a trigger laser to a system exhibiting periodic SR 
provides a deterministic seed, 
which accelerates coherence development and thereby 
may shorten the SR period 
while suppressing its fluctuation 
(see the red solid line in Fig. \ref{fig:conceptual} (b)).
This concept can be referred to as ``triggered periodic SR," 
which represents a step toward enhanced temporal control 
of coherence development in solid-state systems.
If periodic SR can be stabilized, 
it may serve as a practical light source 
producing nanosecond optical pulses.
Our previous measurements showed that 
the linewidth of SR pulses emitted from an Er:YSO crystal was 
much narrower than the inhomogeneous broadening.
Furthermore, precise control of the SR emission timing 
is equivalent to controlling the moment 
when the system reaches maximum coherence.
Performing other quantum operations at this moment---such as 
laser-driven manipulations---allows 
the effects of collective coherence to be fully utilized 
and enables synchronization with the system's peak coherence.

In this paper, we begin by investigating the effect of a trigger laser, whose wavelength is tuned to the same wavelength as the SR transition, 
on the behavior of periodic SR.
Specifically, we measure the dependence of periodic SR 
on the trigger laser power 
under conditions where periodic SR can occur 
even without the trigger laser.
We extend our previous simulation method 
for periodic SR \cite{Hara-SRsim-2026} 
to incorporate the effect of the trigger laser 
and attempt to reproduce 
the experimentally observed dependence on trigger power.
We then examine the role of the trigger laser 
in conditions where periodic SR does not occur 
in the absence of the trigger laser.
These experiments suggest 
the possibility of controlling SR emission timing 
using a short-duration trigger laser pulse 
and this system can be regarded as a device 
capable of SR pulse emission at desired timing.

\section{Experimental setup}
\label{sec:setup}

\begin{figure*}[t]
\begin{center}
      \includegraphics[width=11cm]{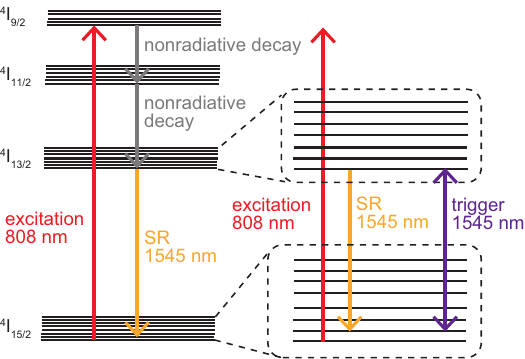}
       \caption{Energy diagram of Er$^{3+}$ ion doped in YSO crystal.
       The dashed rounded square shows enlarged views of 
       the $^{4}$I$_{13/2}$ and $^{4}$I$_{15/2}$ states, respectively.
      }
       \label{fig:setup}
\end{center}
\end{figure*}

Er$^{3+}$ ions doped at site 2 of a YSO crystal, 
which is cooled at 4 K, are used in our experiment.
The number density of the Er$^{3+}$ ions at site 2 
is $5 \times 10^{18}$ cm$^{-3}$ 
according to the nominal concentration value of $0.1 \%$.
Figure \ref{fig:setup} shows the energy diagram.
The difference from the previous experiment \cite{Hara-pSR} is 
the addition of the trigger laser.
The SR pulses with a wavelength of 1545 nm 
are generated in the transition 
from the lowest Stark level of the $^{4}$I$_{13/2}$ state 
to the second lowest Stark level of the $^{4}$I$_{15/2}$ 
ground state.
To trigger the SR pulses, 
a trigger laser with a wavelength of 1545 nm
is injected into the crystal.
The wavelength of the trigger laser is tuned
to the SR transition wavelength.
The excitation and trigger lasers propagate 
along the $b$ axis of the crystal.
Both lasers are loosely focused on the crystal.
The beam diameters ($2 w_{0}$) at the crystal are roughly 
300 $\mu$m (excitation) and 450 $\mu$m (trigger), respectively.
The input excitation laser power is typically 60 mW.
The trigger laser power is different for each measurement 
and ranges from 10 $\mu$W to 9 mW in this paper.
The polarization of the excitation laser 
is parallel to the $\mathbf{D}_{2}$ axis of the crystal.
The polarization of the SR pulses 
in the absence of the trigger laser 
is approximately parallel to the $\mathbf{D}_{1}$ axis.
The excitation laser is turned on 
for 40 ms ($t=0\sim40$ ms) every 200 ms.
The trigger laser is turned on 
for the middle 10 ms ($t=15\sim25$ ms) during the excitation.
The 40 ms excitation data are taken repeatedly.

%

\section{Results and discussion}
\label{sec:results}
\subsection{Change in SR period caused by trigger laser}
\label{sec:trigger1}

\subsubsection{Experimental results}
\label{sec:experiment}

\begin{figure*}[t]
\begin{center}
      \includegraphics[width=15cm]{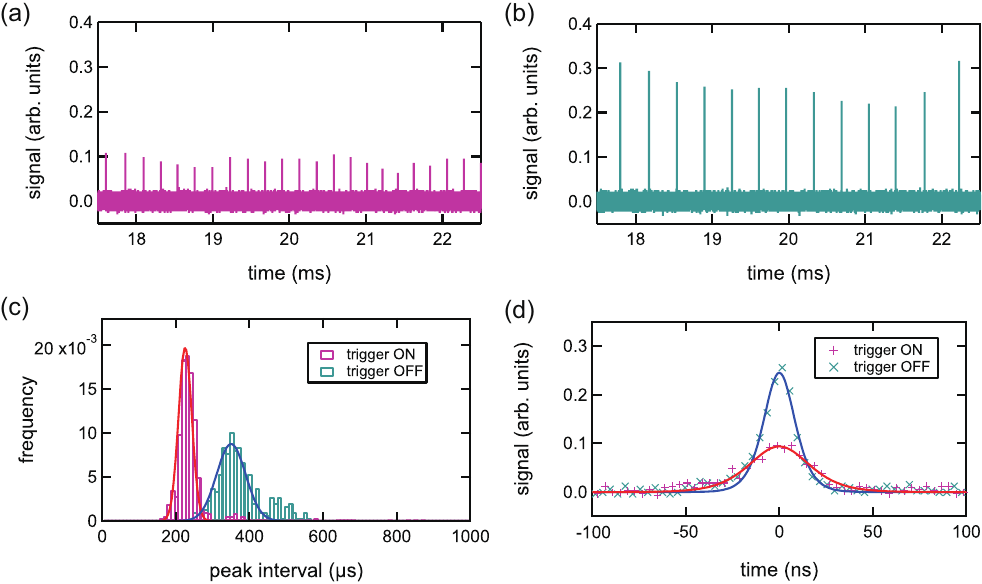}
       \caption{(a) Example of waveform of observed periodic SR pulses with trigger laser irradiation.
       The excitation laser is turned on for 40 ms from $t=0$ 
       and the trigger laser is turned on for 10 ms 
       from $t=15$ ms.
       (b) Example of waveform of observed periodic SR pulses 
       without trigger laser irradiation.
       (c) Histograms of the peak interval 
       between neighboring pulses with (magenta) 
       and without (green) trigger laser irradiation.
       The red and blue lines are the Gaussian fits.
       The data in the time region of $t=15 \sim 25$ ms, 
       during which the trigger laser is irradiated, 
       are used for this plot.
       (d) Examples of observed single SR pulses 
       with (magenta cross) 
       and without (green X) trigger laser irradiation.
       The red and blue solid lines are the fit 
       by sech-squared functions.
       Each peak timing is set as the origin of time.}
       \label{fig:triggeredSR1}
\end{center}
\end{figure*}

First, we observed that the SR period changed 
when the trigger laser was irradiated.
Figure \ref{fig:triggeredSR1} (a) and (b) show 
examples of the waveform of observed periodic SR pulses 
with and without trigger laser irradiation, respectively.
In this case, the trigger laser of 0.5 mW is injected.
The pulses are generated at almost constant time interval 
in both cases.
When triggered, the peak interval decreases 
as we expected in Sec.\ref{sec:introduction}.
Figure \ref{fig:triggeredSR1} (c) shows 
histograms of peak intervals, defined by the time difference 
between neighboring peaks, 
with (magenta) and without (green) trigger laser irradiation.
It can be clearly seen that the peak interval decreases 
when the trigger laser is turned on.
This is consistent with the intuition that 
coherence develops faster when the trigger laser is used as a seed 
rather than spontaneously emitted photons.
The period given by the Gaussian fit is $230 \pm 20$ $\mu$s with trigger 
and $350 \pm 40$ $\mu$s without trigger.
The notation here is (center value $\pm$ $1 \sigma$).
Both the average value of the period and its variance decrease.
This is thought to correspond to the reductions 
of the delay time and its variance in conventional triggered SR.
The decrease in period variability can be interpreted 
as an improvement in the temporal control of coherence development.
In addition to the decrease in period, 
it can be seen that the pulse height also decreases.
Considerations regarding the decrease in pulse height 
will be discussed in Sec. \ref{sec:powdep}, 
which addresses trigger power dependence.
Figure \ref{fig:triggeredSR1} (d) shows 
examples of single SR pulses with (magenta cross) 
and without (green X) trigger laser irradiation.
The FWHM (full-width-half-maximum) pulse durations are 
$37\pm7$ ns with trigger and $19\pm2$ ns without trigger,
respectively.
They are given by the Gaussian fits 
to the histograms of the duration.
In the case of trigger irradiation, 
the increasing of the pulse duration is consistent with 
the decreasing of the peak height, 
and can be understood as conventional SR behavior.
The red and blue solid lines are the fits 
by sech-squared functions.
This functional shape can be seen 
when the coherence develops homogeneously over the target 
\cite{pure-oscillatory-SF-1975} as in our experimental condition.
Other properties regarding frequency and polarization 
are introduced in Appendix (C) and (D).

\subsubsection{Trigger power dependence}
\label{sec:powdep}

\begin{figure*}[t]
\begin{center}
      \includegraphics[width=10cm]{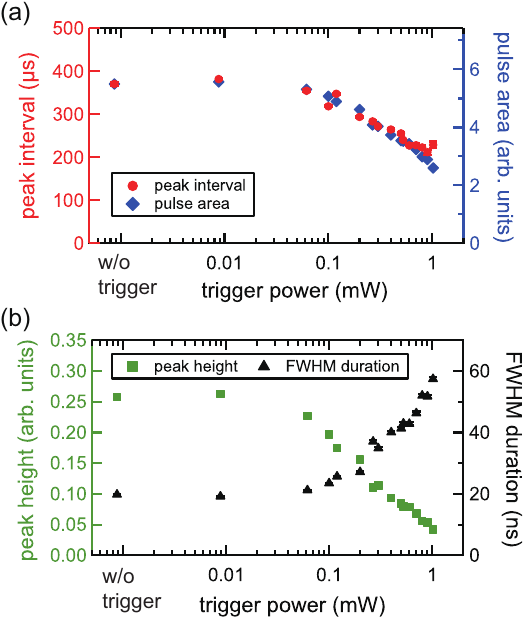}
       \caption{(a) Peak interval (left axis, red circle) and 
       pulse area (right axis, blue diamond) 
       at various trigger laser power.
       (b) Peak height (left axis, green square) 
       and FWHM pulse duration (right axis, black triangle) 
       at various trigger laser power.
       The error bars are the standard errors.
       The leftmost points are the values 
       without the trigger laser.
      }
       \label{fig:powerdep}
\end{center}
\end{figure*}

As shown in Sec. \ref{sec:experiment}, 
the trigger laser affects the behavior of periodic SR.
Next, we 
investigate the effect of trigger laser power on periodic SR.
Figure \ref{fig:powerdep} (a) shows 
the peak interval (left axis, red circle) 
and pulse area (right axis, blue diamond) 
as functions of the trigger laser power.
The leftmost data points correspond to 
the values without the trigger laser.
The vertical axes are normalized such that 
the period and pulse area without the trigger laser 
appear at the same height.
When the trigger power is sufficiently weak, 
the peak interval and pulse area remain close to 
those without trigger laser, 
indicating that the trigger has a negligible effect 
in the weak-power limit.
Above 10 $\mu$W, both period and pulse area decrease 
with increasing trigger power, 
showing that the trigger laser influences the SR behavior 
in the intermediate power range.
Furthermore, the period and the number of emitted SR photons decreases 
while maintaining a proportional relationship with each other.
When the trigger power exceeds 1 mW, SR emission is no longer observed.
This may be due to stimulated emission 
depleting the population of the upper state 
involved in the SR transition, 
preventing the population inversion from reaching 
the threshold required for SR.

The period of SR is shortened by the trigger laser, 
corresponding to the shorter delay time 
in conventional triggered SR \cite{tipping-angle-1979}.
The dependence of the pulse area 
on the trigger laser power can be interpreted 
by consideration of the threshold condition for SR.
For SR to occur, the macroscopic dipole has to develop 
within a decoherence time $T_{2}$ 
and the population inversion has to reach a certain threshold.
That is, $T_{D} < T_{2}$ is required.
The delay time $T_{D}$ is related to the characteristic time of SR, 
which in a simple two-level model is given 
by \cite{Benedict-textbook-1996}
\begin{equation}
T_{\mathrm{R}} = 8 \pi / ( 3 \gamma N_{0} \lambda^{2} L ), 
\end{equation}
through the expression $T_{D} = T_{R} \ln{(N_{0} V)}$ 
\cite{Rydberg-atom-masers-1983}.
Here $N_{0}$ is a number density of Er$^{3+}$ ions related to SR, 
$\lambda$ is a wavelength of SR (1545 nm), 
$L$ is a sample length (6 mm), $\gamma$ is a radiative decay rate, 
and $V$ is an excitation volume.
Defining $f=\ln{(N_{0} V)}$, 
one finds that $f$ typically takes values on the order of 10.
The threshold of the population inversion $N_{\mathrm{th}}$ 
is determined by the above inequality to be 
$N_{\mathrm{th}}=(8 \pi/3 \gamma T_{2}) (V/\lambda^2 L) f$.
As shown in Fig. \ref{fig:conceptual} (a), 
the delay time in the presence of the trigger laser, 
denoted as $T^{\mathrm{trig}}_{D}$, 
is shortened compared to the untriggered case.
This can be expressed as 
$T^{\mathrm{trig}}_{D} = T_{D} / p_{\mathrm{trig}}$, 
where the quantity $p_{\mathrm{trig}} (>1)$ 
is a conceptual parameter that expresses 
how much $T_{D}$ decreases by the trigger.
$p_{\mathrm{trig}}$ is equal to unity 
in the limit of weak trigger laser power, 
and increases as the trigger power becomes stronger.
In the presence of a trigger laser, 
the requirement for SR is expected to be modified to 
$T_{D}^{\mathrm{trig}} < T_{2}$.
The population inversion threshold for SR under the trigger laser, 
$N_{\mathrm{th}}^{\mathrm{trig}}$, is also reduced as 
$N_{\mathrm{th}}^{\mathrm{trig}}=N_{\mathrm{th}} / p_{\mathrm{trig}}$.
This decrease in threshold leads to a corresponding reduction 
in the number of emitted SR photons.
Since the excitation rate remains constant, 
the number of SR photons generated during each cycle 
scales proportionally with the SR period.
This situation is conceptually illustrated 
in Fig. \ref{fig:conceptual} (b, Top).
Thus, a proportional relationship exists 
between the SR period and the number of emitted photons.

Figure \ref{fig:powerdep} (b) shows 
the peak height (left axis, green square) and 
FWHM pulse duration (right axis, black triangle), respectively.
When the trigger power is low, 
both values are comparable to those observed without the trigger laser.
Above 10 $\mu$W, the peak height decreases 
while the FWHM pulse duration increases.
In general, the peak height is proportional to 
the square of the number of SR photons; 
thus, a decrease in peak height indicates 
a reduction in the SR photon number, 
which is consistent with this observation.
This trend, where reduced SR photon number leads to 
a lower peak height and a broader pulse duration, 
is also in agreement with the characteristics of conventional SR.

\subsubsection{Numerical simulation}
\label{sec:simulation}

To gain a quantitative understanding of this phenomenon, 
we perform numerical simulations aimed at reproducing 
the experimentally observed changes in periodic SR behavior 
induced by the trigger laser, 
with a particular focus on its power dependence.

\paragraph{Basic equations}
\label{subsec:basic}

\begin{figure*}[t]
\begin{center}
      \includegraphics[width=10cm]{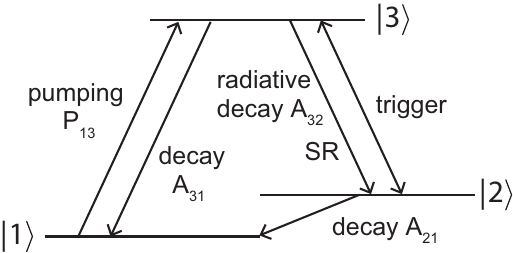}
       \caption{Reduction to extend two-level system in our model.
       The states $\ket{1}$ and $\ket{2}$ correspond to 
       the lowest and second lowest Stark levels of 
       the $^{4}$I$_{15/2}$ ground state, respectively.
       The state $\ket{3}$ corresponds to the lowest Stark level 
       of the $^{4}$I$_{13/2}$ state.
       }
       \label{fig:threelevel}
\end{center}
\end{figure*}

Previously, in Ref. \cite{Hara-SRsim-2026}, 
we constructed a model for periodic SR 
in the absence of the trigger laser.
Here, we will apply such a model 
to consider the case where a trigger laser is added.
The model we will construct is based on 
the Maxwell-Bloch equations for a reduced level system 
shown in Fig. \ref{fig:threelevel}, 
a pair of SR states plus a population reservoir state.
The states $\ket{1}$ and $\ket{2}$ correspond to 
the lowest and the second lowest Stark levels of 
$^{4}$I$_{15/2}$ ground state, 
and $\ket{3}$ corresponds to the lowest Stark level of 
$^{4}$I$_{13/2}$ state, respectively.
A single pumping rate $P_{13}$ is used to represent 
both the excitation to the lowest Stark level of 
the $^{4}$I$_{9/2}$ state and subsequent nonradiative decay 
to $\ket{3}$.
Among the various decay processes from $\ket{3}$, 
only the radiative decay rate $A_{32}$ 
from $\ket{3}$ to $\ket{2}$ plays a role in SR.
Coherence is assumed to develop only between these states.
We collectively represent the other deexcitation processes 
from $\ket{3}$ that do not contribute to SR 
by the incoherent deexcitation rate $A_{31}$ to $\ket{1}$.
The ions in $\ket{2}$ return to $\ket{1}$ 
due to the nonradiative decay of the rate $A_{21}$.

By applying a rotating-wave and slowly varying envelope 
approximations to the Maxwell-Bloch equations 
\cite{Benedict-textbook-1996}, 
the following set of equations can be derived.
\begin{align}
\label{eq:drho11dt}
&\frac{d \rho_{11}}{d t} = -P_{13} \rho_{11} + A_{31} \rho_{33} + A_{21} \rho_{22} , \\
\label{eq:drho22dt}
&\frac{d \rho_{22}}{d t} = \Omega_{s} \rho_{32} + A_{32} \rho_{33} - A_{21} \rho_{22}, \\
\label{eq:drho33dt}
&\frac{d \rho_{33}}{d t} = - \Omega_{s} \rho_{32} + P_{13} \rho_{11} - (A_{31} + A_{32}) \rho_{33}, \\
\label{eq:drho32dt}
&\frac{d \rho_{32}}{d t} = \frac{ \Omega_{s}}{2} ( \rho_{33} - \rho_{22} ) - \gamma_{32} \rho_{32}, \\
\label{eq:Maxwell}
&\frac{\partial \Omega_{s}}{\partial t} = - \kappa \Omega_{s} + \Omega_{0}^{2} \rho_{32} + \Omega_{0} R_{\mathrm{sp}} \rho_{33} + \alpha \kappa_{0} \Omega_{\mathrm{trig}}.
\end{align}
In this paper, 
we incorporate a new effect by the trigger laser 
into the last term of Eq. \ref{eq:Maxwell}, 
which is introduced in Ref. \cite{Rydberg-atom-masers-1983}.
The quantity $\Omega_{\mathrm{trig}}$ is proportional to 
the trigger laser electric field $E_{\mathrm{trig}}$.
The coefficient $\alpha$ represents the degree of influence 
of the $E_{\mathrm{trig}}$ on SR.
It is important to note that among 
the two quantities $\Omega_{s}$ and $\Omega_{\mathrm{trig}}$ 
for the electric field in Eq. (\ref{eq:Maxwell}), 
$\Omega_{s}$ is the variable to be solved in the differential equations 
and $\Omega_{\mathrm{trig}}$ is a constant input parameter.
A detailed explanation of these equations, 
including parameters, is given in Appendix (A).

In our previous theoretical study \cite{Hara-SRsim-2026}, 
we introduced a dynamic modulation of the field decay rate $\kappa$ 
to reproduce the experimentally observed periodic SR.
Following the same approach, we again allow $\kappa$ to switch 
between two values in response to the electric field, 
as described below.
\begin{equation}
\label{eq:kappa}
    \kappa (E_{0}) = \left\{
    \begin{array}{ll}
       \kappa_{0}  &  (|E_{0}| < E_{\mathrm{th}}) \\
       \kappa_{0}/q & (|E_{0}| > E_{\mathrm{th}}), 
    \end{array}
    \right.
\end{equation}
where $E_{\mathrm{th}}$ is the threshold of the electric field.
The parameter $q(>1)$ quantifies the extent of the reduction 
in the field decay rate.



\paragraph{Simulation results}
\label{subsec:simresult}

\begin{figure*}[t!]
\begin{center}
      \includegraphics[width=15cm]{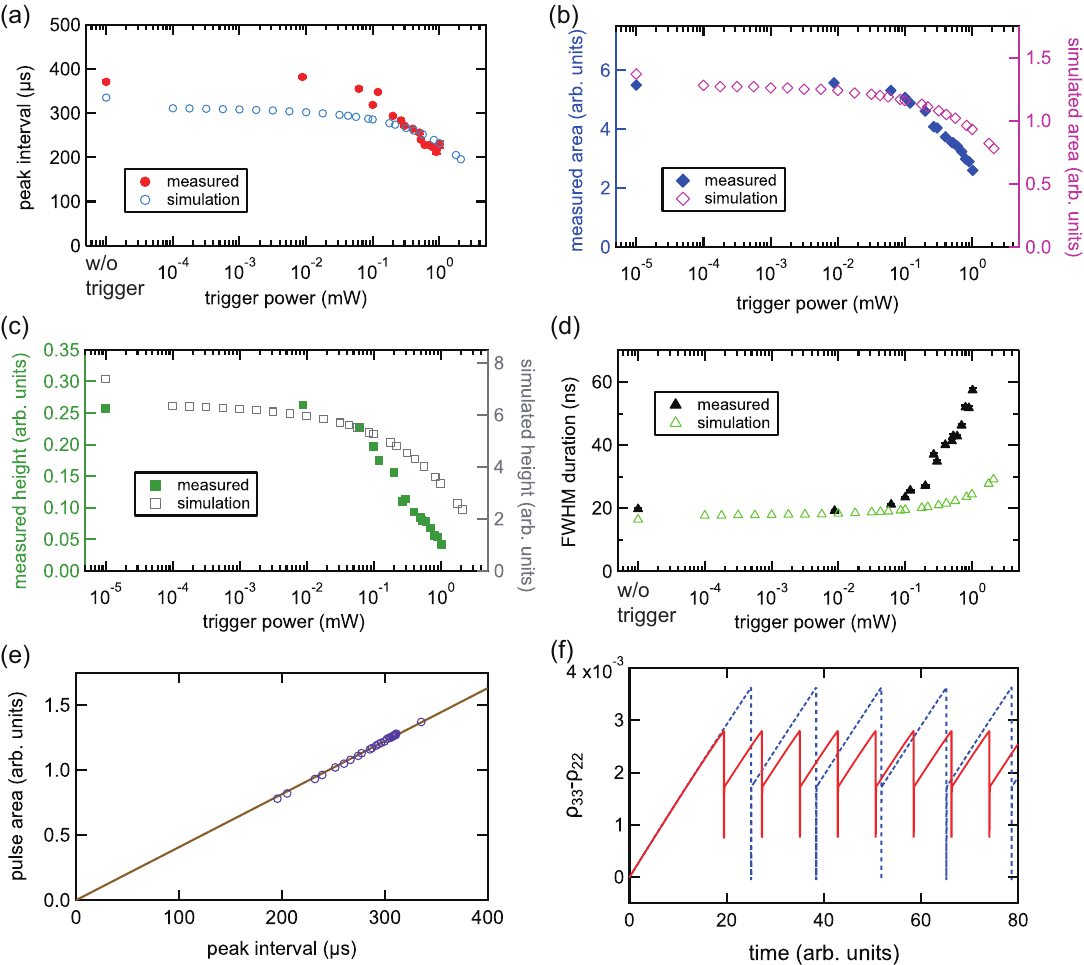}
       \caption{Comparison of the experimental results 
       and numerical simulations.
       The experimental data are shown as filled markers, 
       consistent with Fig. \ref{fig:triggeredSR1}; 
       the simulation results are shown as open markers.
       (a) Peak interval.
       (b) Pulse area.
       (c) Peak height.
       (d) FWHM pulse duration.
       The error bars in the experimental results 
       are the standard errors.
       The leftmost points are the values 
       without the trigger laser.
       Regarding the area and height of the SR pulses, 
       the left axis corresponds to the experimental values 
       and the right axis to the simulation results.
       (e) Pulse area plotted as a function of the period 
       using the simulated values shown in (a) and (b).
       The solid line represents a linear fit 
       constrained to pass through the origin.
       (f) Temporal evolution of population inversion 
       ($\rho_{33} - \rho_{22}$) without (blue dashed) 
       and with (red solid) trigger laser irradiation.}
       \label{fig:simulation}
\end{center}
\end{figure*}

In this section, we present numerical simulation results 
and compare them with the experimental results.
The parameters used for simulations are introduced in Appendix (B).
Figure \ref{fig:simulation} shows comparison of 
the trigger power dependence of 
(a) peak interval, (b) pulse area, (c) peak height, 
and (d) FWHM pulse duration between the experiment and simulation.
The experimental data are shown as filled markers, 
consistent with Fig. \ref{fig:triggeredSR1}, 
while the simulation results are shown as open markers.
The leftmost points are the values without the trigger laser.
Although the peak interval and FWHM duration can be directly compared, 
the pulse area and peak height are not directly comparable 
between the experimental and simulated results.
Therefore, the experimental values are plotted on the left axis, 
and the simulation results on the right.

For the simulation, some of the parameters are adjusted 
to reproduce the experimental results.
We use a parameter set 
$(P_{13}, A_{21}, A_{31}, A_{32}, \gamma_{32})=(6,2.5\times10^{6},104,4,10^{8})$ Hz, 
$(q,E_{\mathrm{th}})=(2,5\times10^{-6}\times(\sqrt{3} \hbar \Omega_{0}/d_{32}))$, 
and $\alpha=3.8\times10^{-2}$.
First of all, the pumping rate $P_{13}=6$ Hz is adjusted to reproduce 
the experimental period in the absence of the trigger laser.
When the actual value of 200 Hz is used in the simulation, 
the resulting period becomes more than an order of magnitude shorter 
than that observed experimentally.
As mentioned in our previous paper \cite{Hara-SRsim-2026}, 
the effective value of $P_{13}$ may reasonably be lower 
if relaxation pathways bypassing $\ket{3}$ exist.
Such pathways would reduce population accumulation 
and thus decrease the effective pumping rate to $\ket{3}$.
The parameter $q=2$ for $\kappa$ modulation is taken 
from our previous study \cite{Hara-SRsim-2026}.
Varying the value of $E_{\mathrm{th}}$ has a small effect 
on the simulation results.
However, both $E_{\mathrm{th}}$ and $\alpha$ need to be carefully 
adjusted to reproduce periodic SR, 
even under strong trigger laser conditions.
The values of $E_{\mathrm{th}}$ and $\alpha$ used in this simulation 
were determined through iterative parameter tuning.

We closely examine the simulation results 
and compare them with experimental observations.
Our simulation reproduces the decrease in 
the peak interval, pulse area, and pulse height 
with increasing trigger laser power, consistent with experiment.
The absolute values obtained in both experiment and simulation 
are in good agreement.
The FWHM duration increases with the strong trigger laser power, 
again consistent with the experiment.
In the weak trigger power range, 
the trigger laser has a negligible effect.
In the strong power range above a few mW, SR emission is not reproduced.
While these overall trends agree reasonably well 
between experiment and simulation, 
the absolute value of the FWHM duration of SR pulses 
is larger in the experiment than in the simulation.
The origin of this discrepancy remains unclear.

Similar to the experimental results, 
the period and pulse area in the simulation 
also show a proportional relationship 
as shown in Fig. \ref{fig:simulation} (e).
The solid line is a linear fit 
constrained to pass through the origin.
This proportional relationship can be understood 
by the constant excitation rate 
and the decrease of the threshold for SR generation.
Figure \ref{fig:simulation} (f) shows temporal evolutions of 
population inversion ($\rho_{33} - \rho_{22}$) 
without (blue dashed) and with (red solid) trigger laser irradiation.
The rate of increase in the population inversion is the same 
due to the same excitation rate.
In the absence of the trigger laser, 
when the population inversion reaches a certain threshold, 
it rapidly decreases due to SR generation.
Subsequently, the population inversion increases rapidly 
as ions in $\ket{2}$ deexcite to $\ket{1}$ 
with a lifetime of $\sim$ 400 ns.
The constant increase of the population inversion begins again 
due to the accumulation of $\ket{3}$ population.
Our simulation shows that the threshold of SR generation decreases 
when the trigger laser is present.
This result explains why both the period and pulse area decrease 
while maintaining a proportional relationship.

The influence of the trigger laser, 
quantified by the parameter $\alpha$ ($\sim 10^{-2}$), is very weak.
This may be attributed to the limited spatial overlap 
between the SR mode and the trigger laser, 
since the propagation direction of SR differs from 
those of the excitation and trigger lasers 
despite optimized alignment of the trigger laser.
Another possible reason is poor frequency mode matching 
between SR and the trigger laser.
The trigger laser frequency cannot be adjusted 
with a precision better than approximately 100 MHz 
and its typical linewidth is on the order of 1 MHz, 
while the linewidth of SR is roughly 100 MHz, 
and individual SR pulses may have even narrower linewidths.
As a result, the frequencies of SR and the trigger laser 
may not match exactly.

The $\kappa$ modulation introduced in Ref. \cite{Hara-SRsim-2026}, 
derived from a theoretical model but not yet experimentally verified, 
yields reasonably accurate results.
This result further supports the validity of 
employing dynamically varying parameters in modeling the system.

\subsection{Trigger-enabled SR under optical excitation}
\label{sec:trigger2}

\begin{figure*}[t!]
\begin{center}
       \includegraphics[width=15cm]{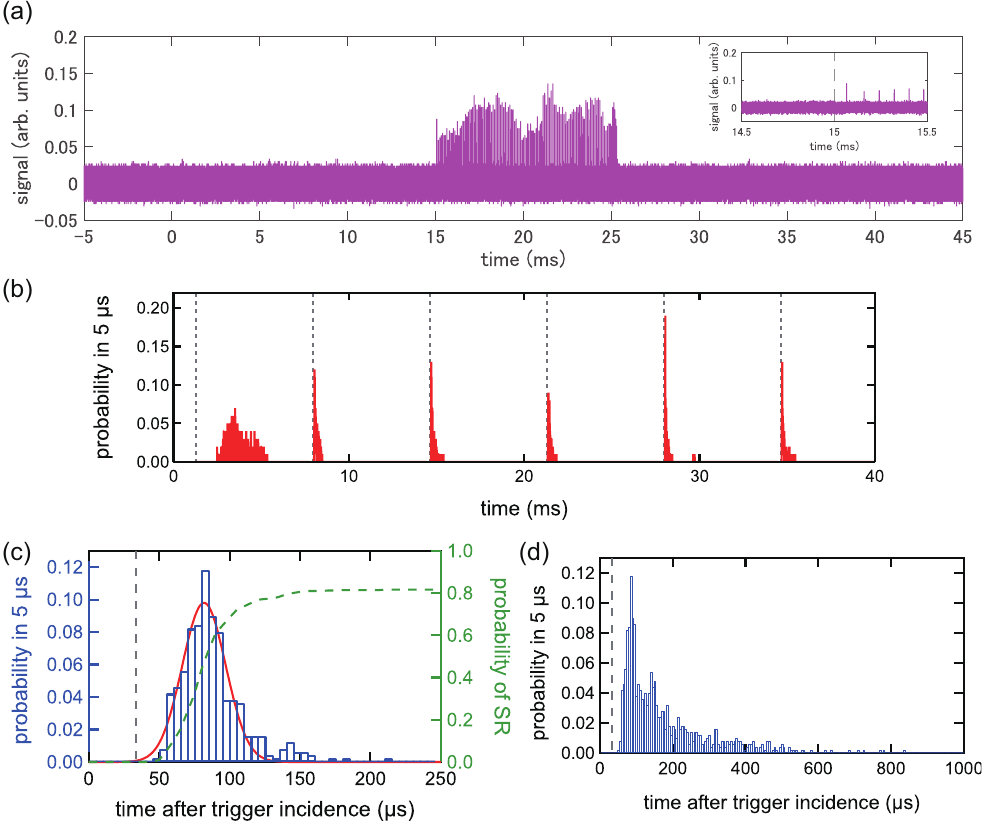}
       \caption{(a) Example waveform of observed periodic SR pulses.
       The excitation laser is turned on for 40 ms from $t=0$, 
       and the trigger laser is turned on for 10 ms from $t=15$ ms.
       No SR is generated in the absence of the trigger laser.
       The inset is an enlarged view of the SR pulse behavior 
       around the onset of the trigger irradiation 
       (gray dashed line at $t=15$ ms).
       (b) Probability of SR occurrence within 5 $\mu$s intervals.
       The trigger laser, with a duration of 33 $\mu$s, is 
       applied at the timing indicated by the dashed lines.
       (c) Probability (in 5 $\mu$s bins) of the timing 
       of the first SR pulse for each trigger laser irradiation.
       The blue bars (left axis) represent the measured data, 
       and the red line shows a Gaussian fit.
       The green dashed line (right axis) represents 
       the cumulative of the blue bars.
       (d) Probability (in 5 $\mu$s bins) of the timing 
       of the all SR pulse for each trigger laser irradiation.
       }
       \label{fig:short}
\end{center}
\end{figure*}

We performed similar experiments under conditions 
where no SR is generated with the excitation laser alone.
This condition could be achieved 
by changing the detuning of the excitation laser by 1 GHz.
The change in laser detuning affects 
the excitation rate and decoherence due to crystal temperature, 
resulting in changes in the behavior of periodic SR.
Under the achieved condition, 
the population inversion may not reach the SR threshold 
with the excitation laser alone, 
and a reduction of the threshold by the trigger laser is required.
While the same time sequence as in Sec. \ref{sec:trigger1} 
was used, the trigger laser power in this experiment was 9 mW.
Figure \ref{fig:short} (a) shows an example of the observed waveform.
The excitation laser is turned on for 40 ms from $t=0$, 
and the trigger laser is turned on for 10 ms from $t=15$ ms.
As noted above, SR does not occur without the trigger laser.
Periodic SR occurs only while the trigger laser is turned on.
The inset shows the SR pulses around the onset of 
the trigger irradiation indicated by the gray dashed line at $t=15$ ms.
Strictly speaking, 
SR pulses do not occur immediately after the trigger pulse, 
but rather starts several tens of microseconds 
after the irradiation begins.

As an application of the above result, 
we considered that this system can be regarded as 
a device capable of generating SR pulses at desired timings.
Based on this idea, we thought that the timing of SR generation 
might be controllable by short trigger laser irradiation, 
and performed the following experiment.
The excitation laser is turned on for 40 ms from $t=0$.
The trigger laser with a duration of 33 $\mu$s is irradiated 
at the timing indicated by the dashed line 
in Fig. \ref{fig:short} (b) with a 6.7 ms period (150 Hz). 
Figure \ref{fig:short} (b) shows 
the probability of SR occurrence in 5 $\mu$s.
The trigger laser is irradiated six times 
during the 40 ms excitation.
The first irradiation does not generate any SRs.
It is likely because the excitation is insufficient 
and the population inversion has not yet reached the SR threshold.
SRs occur between the first and second trigger irradiation, 
but the time of their occurrence varies widely.
These SRs are generated even in the absence of the trigger laser.
Even under experimental conditions 
nominally identical to those of Fig. \ref{fig:short} (a), 
SR between the first and second trigger pulses 
appeared in the condition of Fig. \ref{fig:short} (b), 
and the precise factors responsible for this behavior 
are not yet fully understood.
For the second and subsequent irradiations, 
SRs occur immediately after the trigger laser irradiation.

The blue bars in Fig. \ref{fig:short} (c) represent 
the probability of occurrence of the ``first" SR pulse in 5 $\mu$s 
for each trigger laser irradiation.
The horizontal axis is the time after the trigger laser is turned on.
The trigger laser is turned off at the dashed line ($t=$33 $\mu$s).
The rise and fall times of the trigger laser are typically around 100 ns 
and can be considered instantaneous switching in this plot.
The data for the second to sixth irradiations 
in Fig. \ref{fig:short} (b) 
are used for this analysis.
The same measurement was repeated 100 times, 
therefore each bar is the average probability 
for 500 trigger irradiations.
The red line is the Gaussian fit.
Interestingly, SR does not occur 
immediately after trigger laser irradiation; 
instead, the first SR pulse is generated on average 82 $\mu$s 
after the trigger, with a standard deviation of 15 $\mu$s.
This result indicates that, although it is not possible to 
synchronize SR generation exactly with the trigger, 
its timing can be controlled with a precision 
corresponding to a standard deviation of 15 $\mu$s.
As can be seen, SR tends to occur more prominently 
after the trigger laser is turned off.
This behavior may be related to 
improved coherence in the absence of 
laser-induced heating of the crystal.
Alternatively, stimulated emission 
during the trigger irradiation could reduce 
the population of the upper SR state, 
thereby preventing the system from reaching the SR threshold.
However, the underlying mechanism responsible for this tendency 
remains unclear.
The green dashed line represents the cumulative sum of 
the probabilities shown by the blue bars.
This system can be regarded as a device 
that generates SR pulses with an 80 $\%$ probability 
in response to trigger irradiation.
This probability increases 
with stronger or longer trigger laser irradiation.

On the other hand, there is no strict one-to-one correspondence 
between the trigger laser and the SR pulse; 
multiple SR pulses are generated for each trigger irradiation.
Figure \ref{fig:short} (d) shows the probability of occurrence 
of the ``all" SR pulses in 5 $\mu$s for each trigger.
After the trigger irradiation, 
the probability of SR reaches its peak at approximately 80 $\mu$s.
Subsequently, the SR probability gradually decreases, 
and by around 600 $\mu$s, it becomes almost negligible.
A small secondary peak appears at about 140 $\mu$s, 
which is considered to correspond to the second burst of 
periodic SR following the first peak.
See Appendix (E) for more details.


\section{Summary}
\label{sec: Summary}

To achieve more controllable coherence development, 
we investigated the effects of a trigger laser on periodic SR.
As expected from conventional triggered SR, 
both the period and its variance decreased 
upon irradiation with the trigger laser, 
indicating enhanced controllability of coherence development.
In measurements of trigger power dependence, 
although both the period and the number of SR photons decrease 
with increasing the trigger laser power, 
they remain proportional to each other.
This behavior can be understood 
by the decrease in the SR threshold under the constant excitation rate, 
which was also confirmed by numerical simulations.
The fact that 
the field decay rate modulation 
introduced in Ref. \cite{Hara-SRsim-2026} 
is also effective in the presence of a trigger laser 
supports the validity of our model.
Furthermore, we successfully generated SR using the trigger laser 
in situations where periodic SR does not occur 
with an excitation laser alone.
This technique allowed us to control the timing of SR generation 
through irradiation with a short trigger laser pulse 
and yielded a device capable of generating SR pulses at desired timing.
Triggered periodic SR presents potential as a light source 
that generates narrow linewidth optical pulses 
at nearly constant intervals 
composed of a cooled crystal and two CW lasers.
The ability to control and stabilize the periodicity of SR pulses 
may enable applications in quantum information processing, 
quantum sensing, and fundamental physics, 
where additional manipulation can be performed 
by utilizing a maximum coherence.


\section*{Acknowledgments}
This work was supported by JSPS KAKENHI 
(Grants No. JP19H00686, No. JP20H00161, No. JP21H01112, 
No. JP25K00940, JP25K01027, and JP25K07337) 
and the Korea Research Foundation (Grant No. 2020R1A2C3009299).

\section*{Appendix}

\subsection*{(A) Detailed explanation of equations for numerical simulation}

In the main text, the following set of equations are used 
for numerical simulation.
\begin{align}
\tag{\ref{eq:drho11dt}}
\label{eq_app:drho11dt}
&\frac{d \rho_{11}}{d t} = -P_{13} \rho_{11} + A_{31} \rho_{33} + A_{21} \rho_{22} , \\
\tag{\ref{eq:drho22dt}}
\label{eq_app:drho22dt}
&\frac{d \rho_{22}}{d t} = \Omega_{s} \rho_{32} + A_{32} \rho_{33} - A_{21} \rho_{22}, \\
\tag{\ref{eq:drho33dt}}
\label{eq_app:drho33dt}
&\frac{d \rho_{33}}{d t} = - \Omega_{s} \rho_{32} + P_{13} \rho_{11} - (A_{31} + A_{32}) \rho_{33}, \\
\tag{\ref{eq:drho32dt}}
\label{eq_app:drho32dt}
&\frac{d \rho_{32}}{d t} = \frac{ \Omega_{s}}{2} ( \rho_{33} - \rho_{22} ) - \gamma_{32} \rho_{32}, \\
\tag{\ref{eq:Maxwell}}
\label{eq_app:Maxwell}
&\frac{\partial \Omega_{s}}{\partial t} = - \kappa \Omega_{s} + \Omega_{0}^{2} \rho_{32} + \Omega_{0} R_{\mathrm{sp}} \rho_{33} + \alpha \kappa_{0} \Omega_{\mathrm{trig}}.
\end{align}
A detailed explanation of these equations is given below.
Here, $\rho_{ij}$ are the density matrix elements 
of $\ket{i}$ and $\ket{j}$.
The Rabi frequency is defined as 
$\Omega_{s} \equiv i d_{32} E_{0} / \sqrt{3} \hbar$, 
where $d_{32}$ is the transition dipole moment 
between $\ket{3}$ and $\ket{2}$, 
and $E_{0}$ is the slowly varying envelope of the electric field 
due to the SR radiation.
Equation (\ref{eq_app:drho32dt}) 
describes the temporal evolution of the coherence, 
including decoherence at the rate $\gamma_{32}$. 
Equation (\ref{eq_app:Maxwell}), derived from the Maxwell equation, 
describes the time evolution of the SR electric field.
The emission of SR radiation from the crystal 
is characterized by a field decay rate 
$\kappa = \kappa_{0} \equiv c / n_{0} L$, 
where $c$ is the speed of light and $n_{0}=1.8$ is 
the refractive index of the crystal at the SR wavelength.
We call $R_{\mathrm{sp}}$ the coherence trigger rate, 
which represents the contribution of spontaneous emission 
to the SR electric field.
The frequency $\Omega_{0}$, defined as 
$\Omega_{0} \equiv \sqrt{N_{0} d_{32}^{2} \omega_{32}/ 3 \epsilon_{0} \hbar}$, 
characterizes the time scale of the system.
Here $N_{0}$ is the number density of Er$^{3+}$ ions at site 2 
and $\omega_{32}$ is the angular frequency 
between $\ket{3}$ and $\ket{2}$.
A new effect by the trigger laser is incorporated 
into the last term of Eq. \ref{eq_app:Maxwell}.
The quantity $\Omega_{\mathrm{trig}}$ is proportional to 
the trigger laser electric field $E_{\mathrm{trig}}$, defined as $\Omega_{\mathrm{trig}} \equiv i d_{32} E_{\mathrm{trig}} / \sqrt{3} \hbar$.
The coefficient $\alpha$ represents the degree of influence 
of $E_{\mathrm{trig}}$ on SR.
There are several factors that determine $\alpha$, 
one of which is the spatial overlap 
between the excitation and trigger lasers.
In an ideal case, $\alpha = 1$. 
Let us consider the case in which 
the incident trigger laser passes through the crystal 
without interacting with any ions.
In this situation, the time derivative on the left-hand side, 
as well as the second and third terms on the right-hand side 
of the equation, become zero.
As a result, the $\Omega_{s}$ becomes equal to $\Omega_{\mathrm{trig}}$.
It also becomes evident that the coefficient of $\Omega_{\mathrm{trig}}$ is $+\kappa_{0}$ \cite{Rydberg-atom-masers-1983}.

\subsection*{(B) Parameters in simulations}
\label{subsec:param}

\begin{table*}[t]
\caption{Simulation parameters.
The upper part is related to the rates.
The lower part shows the parameters 
related to the trigger laser irradiation and $\kappa$ modulation.
Here, the uncertainty does not mean $1 \sigma$, but is only a guide.
``factor $U$" represents the uncertainty from factor $1/U$ to $U$.}
\label{tab:param}
\begin{tabular}{cccc}
\addlinespace[1.5mm]
\hline
\addlinespace[1.5mm]
symbol & actual value & measure of uncertainty & values used in this paper \\ 
\addlinespace[1.5mm]
\hline
\addlinespace[1.5mm]
$P_{13}$ & 200 Hz & $\pm 100$ Hz & 6 Hz \\ 
$A_{32}$ & 4 Hz & factor 10 & 4 Hz \\ 
$A_{31}$ & 100 Hz & $70 \leq A_{31} \leq 110$ Hz & 104 Hz \\
$A_{21}$ & $2.5 \times 10^{6}$ Hz & factor 3 & $2.5 \times 10^{6}$ Hz \\
$\gamma_{32}$ & $10^{7}$ Hz & factor 10 & $10^{8}$ Hz \\
$R_{\mathrm{sp}}$ &  &  & $10^{-30} \times \Omega_{0}$ \\
\addlinespace[1.5mm]
\hline
\addlinespace[1.5mm]
$\alpha$ & & & $3.8 \times 10^{-2}$ \\
$q$ & & & 2  \\
$E_{\mathrm{th}}$ & & & $5\times10^{-6} \times (\sqrt{3} \hbar \Omega_{0} /d_{32})$ \\
\addlinespace[1.5mm]
\hline
\end{tabular}
\end{table*}

We briefly introduce the parameters used in the simulation.
They are summarized in Table \ref{tab:param}.
The upper part is related to the rates.
The actual values and measure of uncertainties are same as 
those in Ref. \cite{Hara-SRsim-2026}.
The lower part shows the parameters related to 
the trigger irradiation and $\kappa$ modulation.
The right column shows the values used in this simulation.
The value of $\gamma_{32}$ is adjusted within the uncertainty range 
while $P_{13}$ is adjusted beyond the uncertainty range 
so that the period in the absence of the trigger laser 
is comparable to the experimental value.
This adjustment can be justified 
by the consideration of other relaxation pathways bypassing $\ket{3}$.
The quantity $q$ is the same as that in Ref. \cite{Hara-SRsim-2026}.
The values of $\alpha$ and $E_{\mathrm{th}}$ are chosen 
through trial and error, as described in Sec. \ref{subsec:simresult}.
In the numerical simulation, the dependence of each parameter, 
such as period, is investigated 
by changing the trigger laser power $P_{\mathrm{trig}}$.
$\Omega_{\mathrm{trig}}$, as a function of $P_{\mathrm{trig}}$, 
is evaluated from the relation 
$I_{\mathrm{trig}}=\frac{1}{2}c \epsilon_{0} |E_{\mathrm{trig}}|^{2}$ 
with the trigger laser intensity $I_{\mathrm{trig}}$.
For simplicity, we assume a flat beam of radius $w_{0}$ 
instead of a Gaussian beam when determining $I_{\mathrm{trig}}$.

\subsection*{(C) Trigger frequency dependence}

\begin{figure*}[t!]
\begin{center}
      \includegraphics[width=16cm]{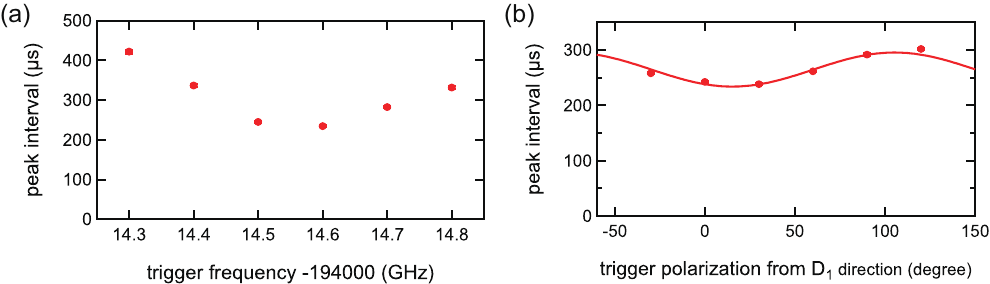}
       \caption{(a) Dependence of peak interval 
       on the trigger laser frequency.
       The red circles and error bars represent 
       the averages and standard errors, respectively.
       (b) Dependence of peak interval on the trigger laser polarization.
       The horizontal axis is the angle of 
       the trigger laser polarization from the $\mathbf{D}_{1}$ axis.
       The red circles and error bars represent 
       the averages and standard errors, respectively.
       The solid line is the fit by a sine function 
       with a fixed period of $180^{\circ}$.}
       \label{fig:trigfreqpol}
\end{center}
\end{figure*}

In the main text, 
we focus on how the trigger laser influences the SR period.
In Appendix (C) and (D), 
we present additional properties of periodic SR. 
The results shown below were obtained 
under the experimental condition described in Sec. \ref{sec:trigger1}.

We varied the trigger laser frequency 
to investigate how wide the frequency range 
affects the behavior of periodic SR.
Figure \ref{fig:trigfreqpol} (a) shows the resulting dependence 
of the peak interval on the trigger laser frequency.
Within the 100 MHz range, 
the peak interval is consistently shorter than 
that in the absence of a trigger laser.
This width is much narrower than the typical inhomogeneous broadening 
in this crystal ($\sim$ 1 GHz) 
and is comparable to the linewidth of SR 
measured in our experiment \cite{Hara-pSR}.
The frequency in Fig. \ref{fig:trigfreqpol} (a) 
is around the SR transition, 
but absolute zero-detuning point 
(\textit{i.e} exact resonance condition) remains uncertain 
due to the wavemeter accuracy 
(WS6-200IR II; HighFinesse, absolute accuracy 120 MHz).
The effective trigger frequency range of approximately 100 MHz is consistent with the intuitive expectation 
that the influence on periodic SR is maximized 
when the trigger frequency matches the SR frequency.

\subsection*{(D) Trigger polarization dependence}

We investigated the influence of trigger laser polarization 
on the periodic SR behavior.
Figure \ref{fig:trigfreqpol} (b) shows 
the dependence of the peak interval 
on the polarization angle of the trigger laser.
The horizontal axis represents the angle from the $\mathbf{D}_{1}$ axis, 
which is approximately aligned with SR polarization 
in the absence of the trigger laser.
The red circles and error bars indicate 
the average values and standard errors, respectively.
The results show that the peak interval becomes shortest 
when the trigger laser polarization is parallel to the SR polarization.
Conversely, the effect on the periodic SR is reduced 
when the trigger laser polarization is orthogonal to the SR polarization.
This polarization dependence is consistent with physical expectations.
The solid line represents a sine-function fit 
with a fixed period of $180^{\circ}$.
The angle at which the minimum peak interval occurs 
is approximately $+15^{\circ}$.
The polarization of the SR is also tilted by a similar angle, 
and the most effective trigger polarization nearly coincides 
with the SR polarization.
There is no particular reason to expect that 
the SR polarization should be strictly aligned with 
the $\mathbf{D}_{1}$ axis.
The $+15^{\circ}$ deviation may include a contribution from 
a slight misalignment 
of the $\mathbf{D}_{1}$ axis relative to the vertical direction 
due to the mounting of the crystal in the copper holder. 

The dependence of the period on the trigger laser polarization 
under the experimental condition described in Sec. \ref{sec:trigger1} 
is consistent with intuition.
In contrast, under the conditions in Sec. \ref{sec:trigger2}, 
SR pulses are generated regardless of the trigger laser polarization, 
even though no SR is generated with the excitation laser alone.
In this case, the polarization of the SR pulses 
is also parallel to the $\mathbf{D}_{1}$ axis.
Such experimental observations, which are counterintuitive 
with respect to the polarization of the trigger laser and SR pulses, 
are not taken into account in the simulations 
presented in Sec. \ref{sec:simulation}.
The underlying mechanism responsible for 
this counterintuitive behavior remains unclear 
and will be the subject of future investigation.

\subsection*{(E) Multiple SR pulses for Short Trigger Irradiation}

\begin{figure*}[t]
\begin{center}
      \includegraphics[width=10cm]{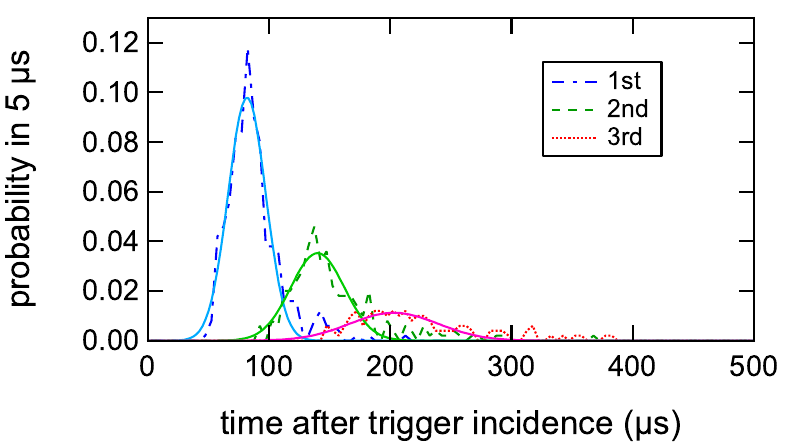}
       \caption{Probabilities (in 5 $\mu$s bins) of the timing of 
       the first (blue dot-dashed line), second (green dashed line), 
       and third (red dotted line) SR pulses 
       for each trigger irradiation.
       The blue dot-dashed line is consistent with 
       Fig. \ref{fig:short} (e).
       The cyan, lightgreen, and magenta lines are Gaussian fits.
       }
       \label{fig:1st2nd3rd}
\end{center}
\end{figure*}

Regarding the short trigger experiment in Sec. \ref{sec:trigger2}, 
we noted that there is no strict one-to-one correspondence 
between the trigger laser and the SR pulse; 
multiple SR pulses are generated for each trigger irradiation.
Figure \ref{fig:short} (d) shows the probability of occurrence 
of the ``all” SR pulses in 5 $\mu$s for each trigger.
Approximately 80 $\mu$s after the trigger irradiation, 
the probability of SR occurrence reaches its peak 
and then gradually decreases.
A small secondary peak appearing at about 140 $\mu$s corresponds 
to the second burst of periodic SR following the first peak.
Figure \ref{fig:1st2nd3rd} shows the probability distributions 
of the timing of the first, second, and third SR pulses 
for each trigger irradiation.
The cyan, lightgreen, and magenta lines are Gaussian fits 
for the corresponding SR pulses.
It can be seen that the probability of the second SR pulse 
reaches its maximum at around 140 $\mu$s.
For the second and third SR pulses, 
the probability of SR occurrence decreases 
and the temporal fluctuation of the emission timing becomes larger.
The probability of the third SR pulse peaks at around 200 $\mu$s.
These SR pulses are generated 
with an average period of approximately 60 $\mu$s.
The first SR pulse exhibits the smallest timing variation 
and is therefore the most precisely controlled.
Note that the SR pulses continue beyond the fourth emission, 
but these subsequent pulses are not shown in Fig. \ref{fig:1st2nd3rd}.


\providecommand{\href}[2]{#2}\begingroup\raggedright\endgroup

\end{document}